\documentstyle[preprint,aps,epsf]{revtex}
\tighten

\let\a=\alpha    
    
 \let\m=\mu \let\n=\nu

\let\C=\Chi

\def\nn{\nonumber} \def\bd{\begin{document}} \def\ed{\end{document}}
\def\ds{\documentstyle} \let\fr=\frac \let\bl=\bigl \let\br=\bigr
\let\Br=\Bigr \let\Bl=\Bigl
\let\bm=\bibitem
\let\na=\nabla
\let\pa=\partial \let\ov=\overline
\let\cal=\mathcal
\newcommand{\be}{\begin{equation}}
\newcommand{\ee}{\end{equation}}
\newcommand{\eeq}{\end{equation}}
\def\ba{\begin{array}}
\def\ea{\end{array}}
\def\ft#1#2{{\textstyle{{\scriptstyle #1}\over {\scriptstyle #2}}}}
\def\fft#1#2{{#1 \over #2}}
\def\del{\partial}
\def\vp{\varphi}
\def\st#1{{\scriptstyle #1}}
\def\sst#1{{\scriptscriptstyle #1}}
\def\bigsq{\mathord{\dalemb{9}{10}\hbox{\hskip1pt}}}
\def\oneone{\rlap 1\mkern4mu{\rm l}}
\def\td{\tilde}
\def\wtd{\widetilde}
\def\ie{\rm i.e.\ }
\def\dalemb#1#2{{\vbox{\hrule height .#2pt
         \hbox{\vrule width.#2pt height#1pt \kern#1pt
                 \vrule width.#2pt}
         \hrule height.#2pt}}}
\def\square{\mathord{\dalemb{6.8}{7}\hbox{\hskip1pt}}}
\def\cramp{\medmuskip = 2mu plus 1mu minus 2mu}
\def\cramper{\medmuskip = 2mu plus 1mu minus 2mu}
\def\crampest{\medmuskip = 1mu plus 1mu minus 1mu}
\def\uncramp{\medmuskip = 4mu plus 2mu minus 4mu}
\newcommand{\ho}[1]{$\, ^{#1}$}
\newcommand{\hoch}[1]{$\, ^{#1}$}
 \newcommand{\bea}{\begin{eqnarray}}
 \newcommand{\eea}{\end{eqnarray}}
\newcommand{\ra}{\rightarrow}
\newcommand{\lra}{\longrightarrow}
\newcommand{\Lra}{\Leftrightarrow}
\newcommand{\bp}{\tilde \beta^\prime}
\newcommand{\tr}{{\rm tr} }
\newcommand{\Tr}{{\rm Tr} }
\def\0{{\sst{(0)}}}
\def\1{{\sst{(1)}}}
\def\2{{\sst{(2)}}}
\def\3{{\sst{(3)}}}
\def\4{{\sst{(4)}}}
\def\5{{\sst{(5)}}}
\def\6{{\sst{(6)}}}
\def\7{{\sst{(7)}}}
\def\8{{\sst{(8)}}}
\def\n{{\sst{(n)}}}
\def\cA{{{\cal A}}}
\def\cF{{{\cal F}}}
\def\tV{\widetilde V}
\def\tW{\widetilde W}
\def\tH{\widetilde H}
\def\tE{\widetilde E}
\def\tF{\widetilde F}
\def\tA{\widetilde A}
\def\im{{{\rm i}}}
\def\jm{{{\rm j}}}
\def\km{{{\rm k}}}
\def\tY{{{\wtd Y}}}
\def\ep{{\epsilon}}
\def\vep{{\varepsilon}}
\def\R{\rlap{\rm I}\mkern3mu{\rm R}}
\def\R{{\mathbf{R}}}
\def\C{{{\mathbf{C}}}}
\def\H{{\mathbf{H}}}
\def\CP{{\mathbf{CP}}}
\def\RP{{\mathbf{RP}}}
\def\Z{{\mathbf{Z}}}
\def\bA{{\mathbf{A}}}
\def\bB{{\mathbf{B}}}
\def\Bbb#1{\mathbf{#1}}
\def\bC{{{\mathbf C}}}
\def\bD{{{\mathbf D}}}
\newcommand{\NP}{Nucl. Phys. }
 \newcommand{\upenn}{Department of Physics and Astronomy\\
 University of Pennsylvania, Philadelphia,  PA 19104, USA}
\newcommand{\itp}{Institute for Theoretical Physics\\
 University of California, Santa Barbara, CA 93106-4030}

\bigskip
\preprint{UPR--998--T, MCTP-02-29, DAMTP-2003-73, CTP TAMU-14/02,
NI02016-MTH, hep-th/0206154}
\bigskip
\begin{document}
\title{Special Holonomy Spaces and M-theory
\footnote{Based on Les Houches 2001 lectures  given by M. Cveti\v c.}
}
\medskip
\author{M. Cveti\v c$^{1,5}$, G.W. Gibbons$^{2}$, H. L\" u$^{3}$ and 
C.N. Pope$^{4,5}$}
\address{$^{1}$\upenn \\ 
 $^2$ DAMTP, Centre for Mathematical Sciences,  Cambridge University\\
Wilberforce Road, Cambridge CB3 OWA, UK\\ 
 $^3$ Michigan Center for Theoretical Physics,
University of Michigan\\ Ann Arbor, MI 48109, USA\\ 
$^4$ Center for Theoretical Physics,
Texas A\&M University, College Station, TX 77843, USA\\ 
$^5$ Isaac Newton Institute for Mathematical Sciences,\\
20 Clarkson Road,  University of Cambridge, 
Cambridge CB3 0EH, UK\\ }
\maketitle
\bigskip
\medskip
\begin{abstract}
We review the construction of regular $p$-brane solutions of M-theory
and string theory with less than maximal supersymmetry whose
transverse spaces have metrics with special holonomy, and where
additional fluxes allow for brane resolutions via transgression terms.
We summarize properties of resolved M2-branes and fractional D2-branes,
whose transverse spaces are Ricci flat eight-dimensional and
seven-dimensional spaces of special holonomy.
Recent developments in the construction of new $G_2$
holonomy spaces are also reviewed.
\end{abstract}


\section{Introduction}
\label{Introduction}

Regular supergravity solutions with less than maximal supersymmetry
may provide viable gravity duals to strongly coupled field theories
with less than maximal supersymmetry. In particular, the regularity of
such solutions at small distances sheds light on confinement and
chiral symmetry breaking in the infrared regime of the dual strongly
coupled field theory \cite{klst}.

We shall briefly review the construction of such regular supergravity
solutions with emphasis on resolved M2-branes of 11-dimensional
supergravity and fractional D2-branes of Type IIA supergravity, which
provide viable gravity duals of strongly coupled three-dimensional
theories with ${\cal N}=2$ and ${\cal N}=1$ supersymmetry.

    This construction has been referred to as a ``resolution via
transgression'' \cite{clp1}.  It involves the replacement of the
standard flat transverse space by a smooth space of special holonomy,
i.e. a Ricci-flat space with fewer covariantly constant spinors.
Furthermore, additional field strength contributions are involved,
which are provided by harmonic forms in the space of special holonomy.
Transgression--Chern-Simons terms modify the equation of motion and/or
Bianchi identity for the original $p$-brane field strength.  In
Section \ref{Resolution} the construction will be reviewed in general
and then applied to resolved M2-branes and D2-branes.

   The explicit construction of such solutions has led to mathematical
developments, for example obtaining harmonic forms for a large class
of special holonomy metrics. As a prototype example we shall review
the construction of the metric and the middle-dimensional forms for
the Stenzel manifolds in $D=2n$ (with $n\ge 2$ integer)
\cite{cglp1,cglp3}. We shall also briefly mention examples of known
$G_2$ holonomy spaces and their associated harmonic forms.  We also
discuss the old as well as the new two-parameter metric with Spin(7)
holonomy \cite{cglpspin7,cglpspin7m} and the associated harmonic
forms.  We shall then briefly summarize the properties of resolved
M2-branes \cite{clp1,cglp3} and fractional D2-branes \cite{clp1,cglp2}
as well as fractional M2-brane whose transverse space is that of the
new Spin(7) holonomy metrics \cite{cglpspin7}.  All these
developments will be reviewed in in Section \ref{Special}.

In the subsequent Section \ref{New} we review the most recent progress
on explicit constructions of $G_2$ holonomy spaces. In particular, we
highlight the construction of general $G_2$ holonomy spaces
whose principal orbits are $S^3$ bundles over $S^3$ and the intriguing
connection of those spaces to a unified description of deformed and
resolved conifolds in six-dimensions.

In Section \ref{Conclusions} we also outline directions of current and
future research, in particular the study of singular $G_2$ holonomy
spaces and their implications for four-dimensional non-Abelian chiral
theories that can arise from a compactification of M-theory on such
classes of special holonomy spaces.

    The work presented in these lectures was initiated in \cite{clp1}
and further pursued in a series of papers that provide both new
technical mathematical results and physics implications for resolved
$p$-brane configurations
\cite{cglp1,cglp3,cglpspin7,cglpspin7m,cglp2}.  Recent progress in the
construction of new special holonomy spaces was initiated in
\cite{cglpspin7,cglpspin7m}, resulting in the first example of
asymptotically locally conical metric with Spin(7)
holonomy. A subsequent series of papers, which appeared after the
lectures had been given, developed these techniques and provided
explicit analyses of classes of cohomogeneity-one $G_2$ holonomy
metrics, with the primary focus on those whose principal orbits are
$S^3$ bundles over $S^3$
\cite{clpmassless,cglp5,bggg,cglp6,Kanno,cglp8,Gukov,cglp10%
,Curio,KannoII,cglp11,brand,cglp12,cglp13,zaslow,cglp14}.
 
\section{Resolution via Transgression}
\label{Resolution}
\subsection{Motivation}

     The AdS$_{D+1}$/CFT$_D$ correspondence \cite{ma,gkp,wi} provides
a quantitative insight into strongly coupled superconformal gauge
theories in $D$ dimensions, by studying the dual supergravity
solutions.  The prototype supergravity dual is the D3-brane of Type
IIB theory, with the classical solution
\bea ds_{10}^2 &=& H^{-1/2}\, dx\cdot dx +
H^{1/2}\, (dr^2 + r^2\,
d\Omega_5^2)\,,\nn\\
F_\5 &=& d^4x\wedge dH^{-1} + {\hat *(d^4x\wedge dH^{-1})}\,,\nn\\
H &=& 1 + \fft{R^4}{r^4}\,\,\, .
 \label{D3N4}
\eea
In the decoupling limit $H = 1 + \fft{R^4}{r^4}
\rightarrow  \  \fft{R^4}{r^4}$ this reduces to $AdS_5\times S^5$,
which provides a  gravitational dual  of  the strongly coupled
${\cal N}=4$ super-Yang-Mills  (SYM) theory.

   Of course, the ultimate goal of this program is to elucidate
strongly coupled YM theory, such as QCD, that has no
supersymmetry. But for the time being important steps have been taken
to obtain viable (regular) gravitational duals of strongly coupled
field theories with less than maximal supersymmetry. In particular,
within this framework we shall shed light on gravity duals of field
theories in $D=\{2,3,4\}$ with ${\cal N}=\{1,2\}$ supersymmetry.

     As a side comment, within $D=5$ ${\cal N}=2$ gauged supergravity
progress has been made (see \cite{becv,ceetal,beda} and references
therein) in constructing domain wall solutions, both with
vector-multiplets {\it and} hyper-multiplets, which lead to smooth
solutions that provide viable gravity duals of $D=4$ ${\cal N}=1$
conformal field theories.  Note however, that often the higher
dimensional interpretation of this approach, and thus a direct
connection to string and M-theory, is not clear.  The aim in these
lectures is to discuss the string and M-theory embeddings of
configurations with less than maximal supersymmetry, and the field
theory interpretation of such gravity duals.

   A procedure for obtaining a supergravity solution with lesser
supersymmetry is to replace the flat transverse 6-dimensional space
$ds_6^2 =dr^2 + r^2\, d\Omega_5^2$ of the D3-brane in (\ref{D3N4})
with a smooth non-compact Ricci-flat space with fewer Killing spinors.
In this case the metric function $H$ still satisfies $ \square H =0$,
but now $\square$ is the Laplacian in the new Ricci-flat transverse
space.  This procedure ensures one has a solution with reduced
supersymmetry; however the solution for $H$ can be singular at the
inner boundary of the transverse space, signifying the appearance of
the (distributed) D3-brane source there.

  A resolution of the singularity (and the removal of the additional
source) can take place if one turns on additional fluxes
(``fractional'' branes). Within the D3-brane context, the Chern-Simons
term of type IIB supergravity modifies the equations of motion:
\bea
dF_\5& = &d{*F_\5} = F_\3^{\rm NS} \wedge F_\3^{\rm RR} =\ft1{2i}
F_\3 \wedge \bar F_\3\,,\nn\\ F_\3 &\equiv& F_\3^{\rm RR} + {\rm i}\,
F_\3^{\rm NS}=m L_\3\, ,
\label{D3cs}
\eea
where $L_\3$ is a complex harmonic self-dual 3-form on the
6-dimensional Ricci-flat space. Depending on the properties of $L_3,$
this mechanism may allow for a smooth and thus viable supergravity
solution. This is precisely the mechanism employed by Klebanov and
Strassler, which in the case of the deformed conifold yields a
supergravity dual of $D=4$ ${\cal N}=1$ SYM theory.  (For related and
follow up work see, for example,
\cite{grpo,manu,gu,pats,bebe,bvflmp,ah,ca,gaetal}.  For earlier work
see, for example, \cite{klwi1,gukl,klne,klts}.)

   In a general context, the resolution via transgression \cite{clp1}
is a consequence of the Chern-Simons-type (transgression) terms that
are ubiquitous in supergravity theories.  Such terms modify the
Bianchi identities and/or equations of motion when additional field
strengths are turned on.  $p$-brane configurations with
$(n+1)$-transverse dimensions, i.e.  with ``magnetic'' field strength
$F_{(n)}$, can have additional field strengths $F_{(p,q)}$ which, via
transgression terms, modify the equations for $F_{(n)}$:
\begin{equation}
dF_{(n)}= F_{(p)}\wedge F_{(q)}\, ; \ \ \  (p+q=n+1)\, .
\label{transgr}
\end{equation}
If the $(n+1)$-dimensional transverse Ricci-flat space admits a harmonic
$p$-form $L_{(p)}$ then the equations of motion are satisfied if one sets $
F_{(p)} =m L_{(p)}$, and by duality $F_{(q)}\sim \m *L_{(p)}\,
$.  Depending on the $L^2$ normalizability properties of $L_{(p)}$, one may
be able to obtain resolved (non-singular) solutions.

\subsection{Resolved M2-brane}

    The transgression term in the 4-form field equation in
11-dimensional supergravity is given by
\be
d{*F_\4} = \ft12 F_\4\wedge F_\4\,,
\end{equation}
and the  modified  M2-brane Ansatz takes the form
\bea
d\hat s_{11}^2 &=& H^{-2/3}\,
dx^\mu\, dx^\nu\, \eta_{\mu\nu} +
H^{1/3}\, ds_8^2\,,\nn\\
 F_\4 &=& d^3x\wedge dH^{-1} + m\, L_\4\,,
\label{m2sol}\eea
where $L_\4$ is a  harmonic self-dual 4-form  in the 8-dimensional
 Ricci-flat
 transverse space.  The  equation for $H$ is then given by
\be
\square H = -\ft1{48} m^2\, L_\4^2\,.\label{m2har}
\end{equation}
For related work see, for example,
\cite{bebe,bebe0,duetal,hata,kbe,hekl}.

\subsection{Resolved D2-brane}

The transgression modification in the 4-form field equation in type
IIA supergravity is
\begin{equation}
d(e^{\ft12\phi}\, {\hat * F_4}) = F_\4\wedge F_\3\, ,
\end{equation}
and the modified D2-brane Ansatz takes the form:
\bea
ds_{10}^2 &=& H^{-5/8}\, dx^\mu\, dx^\nu\, \eta_{\mu\nu} +
H^{3/8}\, ds_7^2\,,\nn\\
F_\4 &=& d^3x\wedge dH^{-1} + m\, L_\4\,,\ \ \
F_\3 = m\, L_\3\,,\ \ \  \phi = \ft14\log H\,,\label{d2sol}
\eea
where $G_\3$ is a harmonic 3-form in the Ricci-flat 7-metric
$ds_7^2$, and $L_\4={*L_\3}$, with $*$ the Hodge dual with respect to
the metric $ds_7^2$.  The function $H$ satisfies
\begin{equation}
 \square H = -\ft16 m^2 L_\3^2\,,\label{d2har}
\end{equation}
where $\square$ denotes the scalar Laplacian with respect to the
transverse 7-metric $ds_7^2$.  Thus the deformed D2-brane solution is
completely determined by the choice of Ricci-flat 7-manifold, and the
harmonic 3-form  supported by it.

\subsection{ Other Examples}
In general the transgression terms  modify field equations or
Bianchi identities as given in (\ref{transgr}), thus allowing
resolved branes with $(n+1)$ transverse dimensions for the
 following  additional examples in M-theory and string theory:
\begin{itemize}
\item (i) D0-brane: $d{*F_\2} = {*F_\4}\wedge F_\3$,
\item (ii) D1-brane: $d{*F}_\3^{\rm RR} = F_\5\wedge F_\3^{\rm NS}$,
\item (iii) D4-brane: $dF_\4 = F_\3\wedge F_\2$,
\item (iv) IIA string: $d{*F_\3} = F_\4\wedge F_\4$,
\item (v) IIB string: $d{*F}_\3^{\rm NS} = F_\5\wedge F_\3^{\rm RR}$,
\item (vi) heterotic 5-brane:  $dF_\3 = F_\2^i\wedge F_\2^i$.
\end{itemize}

In what follows, we shall  focus on resolved M2-branes and briefly
mention fractional D2-branes.  For details of other examples and
their properties, see e.g., \cite{clp1,cglp1,he,heo}.

\section{Special Holonomy Spaces, Harmonic Forms and Resolved Branes}
\label{Special}

The construction of resolved supergravity solutions necessarily
involves the explicit form of the metric on the Ricci-flat special
holonomy spaces. These spaces fall into the following classes:
\begin{itemize}
\item K\"ahler spaces in $D=2n$ dimensions ($n$-integer) with $SU(n)$
holonomy, and two covariantly constant spinors. There are many
examples, with the Stenzel metric on $T^*S^n$ providing a
prototype. They are typically asymptotically conical (AC).

\item Hyper-K\"ahler spaces in $D=4n$ with $Sp(n)$ holonomy, and $n+1$
covariantly constant spinors. Subject to certain technical
assumptions, Calabi's metric on the co-tangent bundle of $\CP^n$ is
the only complete irreducible cohomogeneity one example \cite{dasw}.

\item In $D=7$ there are exceptional $G_2$ holonomy spaces with one
covariantly constant spinor. Until recently only three AC examples
were known \cite{brysal,gibpagpop}, but new metrics have been recently
constructed in \cite{bggg,cglp5,Kanno,cglp8,Gukov,cglp10%
,Curio,KannoII,cglp11,brand,cglp12,cglp13,zaslow,cglp14} and will be
discussed in a separate Section \ref{New}.

\item In $D=8$ there are exceptional Spin(7) holonomy spaces with
one covariantly constant spinor; until recently only one AC example
was known \cite{brysal,gibpagpop}. New metrics were recently
constructed in \cite{cglpspin7,cglpspin7m,Kanno,Gukov,cglp10}.
\end{itemize}

   Here the focus is on a construction of cohomogeneity one spaces
that are typically asymptotic to cones over Einstein spaces. Recent
mathematical developments evolved in two directions: (i)
construction of harmonic forms on known Ricci-flat spaces (see in
particular \cite{cglp1,cglp2}), (ii) construction of new exceptional
holonomy spaces
\cite{cglpspin7,cglpspin7m,cglp5,bggg,Kanno,cglp8,Gukov,cglp10%
,Curio,KannoII,cglp11,brand,cglp12,cglp13,zaslow,cglp14} .  In the following
subsections we illustrate these developments by 
\begin{itemize}

\item Summarizing
results on the construction of harmonic forms on the Stenzel metric
\cite{cglp1}, 

\item Briefly mentioning the results for the old $G_2$
holonomy metrics \cite{clp1,cglp2}, 

\item Presenting results for the new
Spin(7) two-parameter metrics \cite{cglpspin7,cglpspin7m}, and

\item Summarizing the implications of these prototype special holonomy
spaces for resolved M2-branes and D2-branes.

\end{itemize}

  In Section \ref{New} we shall then
summarize the recent new constructions of $G_2$ holonomy spaces and
the implications for M-theory dynamics on such spaces.

\subsection{Harmonic forms for the Stenzel metric}

The Stenzel\cite{st} construction provides a class of complete
non-compact Ricci-flat K\"ahler manifolds, one for each even
dimension, on the co-tangent bundle of the $(n+1)$-sphere, $T^\star
S^{n+1}$.  These are asymptotically conical, with principal orbits
that are described by the coset space $SO(n+2)/SO(n)$, and they have
real dimension $d=2n+2$.

\subsubsection{Stenzel metric}

In the following we summarize the relevant results for the
construction of the Stenzel metric. (For more details see
\cite{cglp1}.)  This construction \cite{cglp1,st} of the Stenzel
metric starts with $L_{AB}$, which are left-invariant 1-forms on the
group manifold $SO(n+2)$. By splitting the index as $A=(1,2,i)$, we
have that $L_{ij}$ are the left-invariant 1-forms for the $SO(n)$
subgroup, and so the 1-forms in the coset $SO(n+2)/SO(n)$ will be
\be
\sigma_i \equiv L_{1i}\,,\qquad \td\sigma_i \equiv L_{2i}\,,\qquad
\nu \equiv L_{12}\,.
\end{equation}
The metric Ansatz takes the form:
\begin{equation}
ds^2 = dt^2 + a^2 \sigma_i^2 + b^2\, \td\sigma_i^2 + c^2\,
\nu^2\,, \label{stenmeth}\end{equation}
where $a$, $b$ and $c$ are functions of the radial coordinate $t$. One
defines Vielbeine
\begin{equation}
e^0=dt\,,\qquad e^i = a\, \sigma_i\,,\qquad e^{\td i} = b\,
\td\sigma_i\,,\qquad e^{\td 0}=c\, \nu\, , \label{vielbein}
\end{equation}
for which one can introduce a holomorphic
tangent-space basis of complex 1-forms $\ep^\a$:
\begin{equation}
\ep^0 \equiv -e^0 + \im\, e^{\td 0}\,,\qquad \ep^i = e^i + \im\,
e^{\td i}\,.\label{complexbasis}
\end{equation}

Defining $a=e^\a$, $b=e^\beta$, $c=e^\gamma$, and introducing the new
coordinate $\eta$ by $a^n\, b^n\, c\, d\eta = dt$, one finds
\cite{cglp1} that the Ricci-flat equations can be obtained from a
Lagrangian $L=T-V$ which can be written as a ``supersymmetric
Lagrangian'': $L=\ft12 g_{ij}\, (d{\a^i}/d\eta)\, (d{\a^j}/d\eta) -
\ft12 g^{ij} \, \fft{\del W}{\del \a^i}\, \fft{\del W}{\del \a^j} \,
.$ The solution of the first-order equations yields the explicit
solution:
\begin{equation}
a^2 = R^{1\over{n+1}}\, \coth r\,; 
b^2 = R^{1\over {n+1}}\, \tanh r\,;
h^2=c^2 =\textstyle{ \fft1{n+1}} R^{-{{n}\over {n+1}}} \sinh^n
(2r)\,,
\end{equation}
where
$
R(r) \equiv  \int_0^r (\sinh 2u)^n\, du\, ,
$
and the radial coordinate  $r$ is introduced as $dt=h\, dr$.

    For each $n$ the result is expressible in relatively simple terms.
For example, 
\be
 R = \sinh^2 r\, ; R =  \ft18(\sinh 4 r- 4r)\, ;
 R = \ft23 (2+\cosh 2r)\sinh^4 r \, ,
\end{equation}
for $n=1,2,3$, respectively.
The case $n=1$ is the Eguchi-Hanson metric \cite{egha},
and  $n=2$ it is the deformed conifold \cite{dloc}.

    As $r$ approaches zero, the metric takes the form
\be
ds^2\sim dr^2+r^2{\tilde \sigma}_i^2 +\sigma_i^2+\nu^2\,,
\end{equation}
which has the structure locally of the product
$\R^{n+1}\times S^{n+1}$,  with $S^{n+1}$ being a  ``bolt.''
As $r$ tends to infinity, the metric becomes
\be
ds^2\sim d\rho^2 +\rho^2\{\textstyle{{n^2\over {(n+1)^2}}}\nu^2+
\textstyle{{n\over {2(n+1)^2}}}(\sigma_i^2 +{\tilde \sigma_i}^2)\}\, ,
\end{equation}
representing a cone over the Einstein space $SO(n+2)/SO(n)$.

\subsubsection{Harmonic middle-dimension $(p,q)$ forms}

An Ansatz compatible with the symmetries of the Stenzel metric is of
the form:
\bea
L_{\sst{(p,q)}} &=& f_1\,\ep_{i_1\cdots i_{q-1}j_1\cdots j_{p}}\,
\bar \ep^0\wedge \bar \ep^{i_1}\wedge\cdots \wedge\bar\ep^{i_{q-1}}\wedge
\ep^{j_1}\wedge\cdots\wedge\ep^{j_{p}}\nn\\
&&+f_2\,\ep_{i_1\cdots i_{p-1}j_1\cdots j_{q}}\,
\ep^0\wedge \ep^{i_1}\wedge\cdots\wedge \ep^{i_{p-1}}\wedge
\bar \ep^{j_1}\wedge\cdots\wedge \bar \ep^{j_q}\,,
\label{middled}
\eea
with $f_1$, $f_2$ being functions of $r$, only.
The harmonicity condition becomes $ dL_{\sst{(p,q)}}=0\, , $ since $
{*L_{\sst{(p,q)}}} = \im^{\, p-q}\, L_{\sst{(p,q)}}\,$.  The functions
$f_1$, $f_2$ are solutions of coupled first-order homogeneous
differential equations, yielding a solution that is finite as $r\to
0$:
\be
f_1 =\, q\, _2F_1\left[ \ft12 p, \ft12 (q+1), \ft12 (p+q) +1;
-(\sinh 2r)^2\right]\, ,
\end{equation}
\be
f_2 = -\, p\, _2F_1\left[ \ft12 q, \ft12 (p+1), \ft12 (p+q) +1;
-(\sinh 2r)^2\right]\, .
\end{equation}
For any specific integers $(p,q)$, these are elementary functions of
$r$.

   For the two special cases of greatest interest, they have the
following properties:
\begin{itemize}
\item $(p,p)$-forms in $4p$-dimensions: $f_1=-f_2={p\over
(\cosh{r})^{2p}}$ with $|L_{(p,p)}|^2={const.\over {(\cosh{r})^{4p}}}$
falls-off fast enough as $r\to\infty$.  This turns out to be the only
$L^2$ normalizable form.

\item $(p+1,p)$-forms in $(4p+2)$-dimensions. As $r\to \infty$:
$|L_{(p+1,p)}|^2\sim {1\over {[\sinh{(2r)}]^{2p}}}$ which is
marginally $L^2$-non-normalizable.
\end{itemize}

   From the viewpoint of physics, the case in $2(n+1)=4$ dimensions with
an $L^2$-normalizable $L_{(1,1)}$-form is precisely the example of the
resolved self-dual string discussed in \cite{clp1}.

  In $2(n+1)=6$ dimensions, the $L_{(2,1)}$-form was constructed in
\cite{klst}, and provides a resolution of the D3-brane. Since
$L_{(2,1)}$ is only {\it marginally non-normalizable} as $r\to\infty$,
the decoupling limit of the space-time does not give an AdS$_5$, but
instead there is a logarithmic modification. In particular, this
modification accounts for a renormalization group running of the
difference of the inverse-squares of the two gauge group couplings in
the dual $SU(N)\times SU(N+M)$ SYM \cite{klne}.

    On the other hand in $2(n+1)=8$ dimensions the $L^2$ normalizable
$L_{(2,2)}$-form supports additional fluxes that resolve the original
$M2$-brane, whose details are given in \cite{cglp1}.

   It turns out that one can construct regular supersymmetric resolved
M2-branes for many other examples of 8-dimensional special holonomy
transverse spaces, such as the original Spin(7) holonomy transverse
space \cite{clp1}, a number of new K\" ahler spaces \cite{clp1,cglp2},
and hyper-K\"ahler spaces \cite{cglp3}.

\subsection{Old $G_2$ holonomy metrics and  their harmonic forms}
\subsubsection{ Resolved cones over $S^2\times S^4$  and $S^2\times \CP^2$}

   The first complete Ricci-flat 7-dimensional metrics of $G_2$
holonomy were obtained in \cite{brysal,gibpagpop}.  There were two types.
The first type comprises two examples of $R^3$
bundles over four-dimensional quaternionic-K\"ahler Einstein base
manifolds $M$.  These spaces are of cohomogeneity one, with principal
orbits that are $S^2$ bundles over $M$ (sometimes referred to as the
twistor space over $M$). For the two examples that arise, $M$ is
$S^4$ or $\CP^2$.  The two $G_2$ manifolds have principal orbits
that are $\CP^3$ ($S^2$ bundle over $S^4$), or the flag manifold
$SU(3)/(U(1)\times U(1))$ ($S^2$ bundle over $\CP^2$),
respectively. These two manifolds are the bundles of self-dual 2-forms
over $S^4$ or $\CP^2$ respectively. They approach $\R^3\times S^4$ or
$\R^3\times \CP^2$ locally near the origin. 

     The derivations for the
two cases, with the principal orbits being $S^2$ bundles either over
$S^4$ or over $\CP^2$, proceed essentially identically.
In the notation of \cite{gibpagpop}, the $G_2$ metrics are given by 
\be d\hat s_7^2 = h^2\, dr^2 + a^2\, (D\mu^i)^2 + b^2\,
ds_4^2\,, \label{g2olrd1}\end{equation}
where $\mu^i$ are coordinates on $\R^3$ subject to $\mu^i\,
\mu^i=1$, and $ds_4^2$ is the metric on $S^4$ or $\CP^2$ scaled to have
$R_{\alpha\beta}=3 g_{\alpha\beta}$.   The 1-forms $A^i$ are
$su(2)$ Yang-Mills instanton potentials, and 
\be 
D\mu^i \equiv d\, \mu^i + \ep_{ijk}\, A^j\, \mu^k\,. 
\end{equation}
The field strengths $J^i\equiv dA^i + \ft12 \ep_{ijk}\, A^j\wedge A^k$
satisfy the algebra of the unit quaternions, $J^i_{\a\gamma}\,
J^j_{\gamma\beta} = -\delta_{ij}\, \delta_{\a\beta} + \ep_{ijk}\,
J^k_{\a\beta}$.  The harmonic 3-form (other than the covariant
constant one), was constructed in \cite{cglp3}: it is smooth and 
$L^2$-normalizable.

\subsubsection {Resolved cone over $S^3\times S^3$}

    The second type of complete 7-dimensional manifold of $G_2$ holonomy
obtained in \cite{brysal,gibpagpop} is again of cohomogeneity one,
with principal orbits that are topologically $S^3\times S^3$.  The manifold
is the spin bundle of $S^3$; near the origin it approaches locally
$\R^4\times S^3$.

  The Ricci-flat metric on the spin bundle of $S^3$ is given by
\cite{brysal,gibpagpop}.
\be 
ds_7^2 = \a^2\, dr^2 + \beta^2\, (\sigma_i - \ft12
\Sigma_i)^2 + \gamma^2\, \Sigma_i^2\,,\label{g2old2} 
\end{equation}
where the functions $\a$, $\beta$ and $\gamma$ are given by
\be 
\a^2 = \Big(1-\fft{1}{r^3}\Big)^{-1}\,,\qquad \beta^2 = \ft19
r^2\,  \Big(1-\fft{1}{r^3}\Big)\,,\qquad \gamma^2 = \ft1{12}
r^2\,. 
\end{equation}
Here $\Sigma_i$ and $\sigma_i$ are the two sets of left-invariant
1-forms on two independent $SU(2)$ group manifolds.  The principal
orbits $r=$constant are therefore $S^3$ bundles over $S^3$.  Since the
bundle is trivial, they are topologically $S^3\times S^3$, although
not with the standard product metric. The radial coordinate runs from
$r=a$ to $r=\infty$.

   This metric admits a regular harmonic 3-form, explicitly constructed
in \cite{clp1}: it is square-integrable at short distance, but gives a
linearly divergent integral at large distance.  The short-distance
square-integrability is enough to give a regular deformed D2-brane
solution, even though $L_{(3)}$ is { not $L^2$-normalizable}.

\subsection{New Spin(7) holonomy metrics and their harmonic forms}
\subsubsection{The old metric and harmonic 4-forms}

   Until recently only one explicit example of a complete non-compact
metric on a Spin(7) holonomy space was known
\cite{brysal,gibpagpop}. The principal orbits are $S^7$, viewed as an
$S^3$ bundle over $S^4$. The solution (\ref{spin7metric}) is
asymptotic to a cone over the ``squashed'' Einstein 7-sphere, and it
approaches $\R^4\times S^4$ locally at short distance (\ie
$r\approx\ell$). The metric is of the form:
\be
ds_8^2 = \Big(1- \fft{\ell^{10/3}}{r^{10/3}}\Big)^{-1} \, dr^2
         + \ft{9}{100}\, r^2\, \Big(1- \fft{\ell^{10/3}}{r^{10/3}}\Big)\,
         h_i^2 + \ft{9}{20} r^2\, d\Omega_4^2\,,\label{spin7metric}
\end{equation}
where $h_i\equiv \sigma_i - A_\1^i\,,$ and the $\sigma_i$ are
left-invariant 1-forms on $SU(2)$, $d\Omega_4^2$ is the metric on the
unit 4-sphere, and $A_\1^i$ is the $SU(2)$ Yang-Mills instanton on
$S^4$.  The $\sigma_i$ can be written in terms of Euler angles as
$\sigma_1= \cos\psi\, d\theta\, + \sin\psi\, \sin\theta\,
d\varphi\,,$\\
$\sigma_2= -\sin\psi\, d\theta\, + \cos\psi\, \sin\theta\,
d\varphi\,,$
$\sigma_3 = d\psi\, + \cos\theta\, d\varphi\,$.
A regular $L^2$ normalizable harmonic 4-form in this metric was obtained
in \cite{clp1}.

\subsubsection{New Spin(7) holonomy metric}

 The generalization that we shall consider involves allowing the
$S^3$ fibers of the previous construction themselves to be
``squashed.'' Namely, the $S^3$ bundle is itself written as a
$U(1)$ bundle over $S^2$ leading to the following ``twice
squashed'' Ansatz:
\be
d\hat s_8^2 = dt^2 + a^2\, (D\mu^i)^2 + b^2\, \sigma^2 + c^2\,
d\Omega_4^2\,,\label{8ans}
\end{equation}
where $a$, $b$ and $c$ are
functions of the radial variable $t$. (The previous Spin(7)
example has $a=b$.) Here
\be
\mu_1 = \sin\theta\,
\sin\psi\,,\qquad \mu_2= \sin\theta\, \cos\psi\,,\qquad \mu_3=
\cos\theta\,,
\end{equation}
 are the $S^2$ coordinates, subject to the
constraint $\mu_i\mu_i=1$, and
\be
D\mu^i\equiv d\mu^i
+\ep_{ijk}\, A_\1^j\, \mu^k\,,\quad \sigma \equiv d\varphi +
\cA_\1\,,\quad \cA_\1 \equiv \cos\theta\, d\psi - \mu^i\,
A_\1^i\, , \label{kkvector}
\end{equation}
where the  field strength ${\cal
F}_{(2)}$ of the
 $U(1)$ potential ${\cal A}_{(1)}$ turns out to be given by:
$
\cF_\2 = \ft1{2} \ep_{ijk}\, \mu^k\, D\mu^i \wedge
D\mu^j - \mu^i\, F_\2^{i}$.

   The Ricci-flatness conditions can be satisfied by solving the
first-order equations coming from a supersymmetric Lagrangian, yielding the
following special solution (for details see
\cite{cglpspin7,cglpspin7m}):
\begin{eqnarray}
ds_8^2 &=& \fft{(r-\ell)^2\,
dr^2}{(r-3\ell)(r+\ell)} + \fft{\ell^2\,
(r-3\ell)(r+\ell)}{(r-\ell)^2}\, \sigma^2\nn\\
&+&
\ft14(r^2-3\ell)(r+\ell)\, (D\mu^i)^2 + \ft12(r^2-\ell^2)\,
d\Omega_4^2\,, \label{sol2}
\end{eqnarray}
The quantity $\ft14[\sigma^2 + (D\mu^i)^2]$ is the metric on the unit
3-sphere, and so in this case we find that the metric smoothly
approaches $\R^4\times S^4$ locally, at small distance ($r\to 3\ell$),
in the same way that it does in the previously-known example.
Therefore it has the same topology as the old Spin(7) holonomy
space. On the other hand, it locally approaches ${\cal M}_7\times S^1$
at large distance.  Here ${\cal M}_7$ denotes the 7-manifold of $G_2$
holonomy on the $\R^3$ bundle over $S^4$
\cite{brysal,gibpagpop}. Asymptotically the new metric behaves like a
circle bundle over an asymptotically conical manifold in which the
length of the $U(1)$ fibers tends to a constant; in other words, it is
ALC.

   If one takes $r$ to be negative, or instead analytically continues the
solution so that $\ell \to -\ell$ (keeping $r$ positive), one gets a
different complete manifold. Thus instead of (\ref{sol2}), the
quantity $\ft14(\sigma^2+(D\mu^i)^2 + d\Omega_4^2)$ is precisely the
metric on the unit 7-sphere, and so as $r$ approaches $\ell$ the
metric $ds_8^2$ smoothly approaches $\R^8$.  At large $r$ the function
$b$, which is the radius in the $U(1)$ direction $\sigma$, approaches
a constant, and so the metric tends to an $S^1$ bundle over a 7-metric
of the form of a cone over ${\Bbb C} {\Bbb P}^3$ ; it has the same
asymptotic form as (\ref{sol2}).  The manifold in this case is
topologically $\R^8$.

   In \cite{cglpspin7,cglpspin7m} the general solution to the
first-order system of equations is obtained, leading to additional
families of regular metrics of Spin(7) holonomy, which are complete
on manifolds $\bB_8^\pm$ that are similar to $\bB_8$. These additional
metrics have a non-trivial integration constant which parameterizes
inequivalent solutions. (For details see \cite{cglpspin7m} and
Appendix A of \cite{cglpspin7}).

     $L^2$ normalizable harmonic 4-forms for the new Spin(7)
8-manifolds were obtained in \cite{cglpspin7}.

\subsection{Applications: resolved M2-branes and D2-branes}
The explicit construction of harmonic 4-forms on 8-dimensional
Ricci flat spaces led to analytic expressions for resolved M2-brane
solutions, while the 3-forms (and dual 4-forms) of $G_2$
holonomy spaces led to analytic expressions for fractional D2-branes.
Their properties, such as supersymmetry conditions, flux integrals and
aspects of the dual field theories, were discussed in
\cite{clp1,cglp1,cglp3,cglp2,hekl}.
 
    Resolved M2-branes on a suitable eight-dimensional space can be
supported by $L^2$-normalizable harmonic forms and thus they are
regular at short distance and have decoupling limits at large distance
that yield $AdS_4$.  They have no conserved additional (fractional)
charges.  The dual 3-dimensional field theory is 
superconformal (with ${\cal N}=1$ or ${\cal N}=2$
supersymmetry), and is in turn perturbed by marginal operators
associated with pseudo-scalar fields \cite{hekl}. On the other hand
the fractional D2-branes
have conserved fractional charges corresponding to D4-branes wrapping the
2-cycles dual to $S^4$ or $\CP^2$ in ${\cal M}_7$, or to NS-NS
5-branes wrapping the 3-cycle dual to $S^3$ in ${\cal M}_7$.

   An interesting application of these new Spin(7) holonomy spaces is
the construction of fractional M2-branes.  After reduction on $S^1$
these give D2-branes with additional fractional magnetic charge
associated with D4-branes wrapping 2-cycles {\it and} D6-branes
wrapping 4-cycles.  The fact that the resolved M2-brane on the new
Spin(7) holonomy space has non-zero fractional charge is a
consequence of the asymptotically locally conical structure of the new
Spin(7) holonomy space.

\section{New $G_2$ Holonomy Metrics}
\label{New}

   Subsequent to these lectures there have been major developments the
construction of new holonomy spaces and the study of their
implications for M-theory dynamics.  In part motivated by the
construction of the new two-parameter Spin(7) holonomy metrics with
ALC structure \cite{cglpspin7,cglpspin7m} (described in the previous
Section \ref{Special}), new constructions of $G_2$ (as well as
Spin(7)) holonomy spaces \cite{cglp5,bggg,Kanno,cglp8,Gukov,cglp10%
,Curio,KannoII,cglp11,brand,cglp12,cglp13,zaslow,cglp14} have been given.
The implications from M-theory on such spaces for the dynamics of the
resulting ${\cal N}=1$, $D=4$ field theory \cite{atwi,ac,atmava} are
attracting considerable attention.  Specifically, it has been proposed
that M-theory compactified on a certain singular seven-dimensional
space with $G_2$ holonomy might be related to an ${\cal N}=1$, $D=4$
gauge theory \cite{wi,atwi,ac,atmava,agva} that has no conformal
symmetry.  The quantum aspects of M-theory dynamics on spaces of $G_2$
holonomy can provide insights into non-perturbative aspects of
four-dimensional ${\cal N}=1$ field theories, such as the preservation
of global symmetries and phase transitions.  For example,
Ref. \cite{atwi} provides an elegant exposition and study of these
phenomena using the three original manifolds of $G_2$ holonomy that were
obtained in \cite{brysal,gibpagpop}.

     One related development in this direction is the discovery of
M3-brane configurations \cite{clpmassless,cglp5}.  These have a flat
4-dimensional world-volume and a transverse space that is a
deformation of the $G_2$ manifold, and with the 4-form field strength
is turned on. They turn out to have zero charge and ADM
mass, leading to naked singularities at small distances.

\subsection{Classification of $G_2$ holonomy spaces with
$S^3 \times S^3$ orbits}
   
   In another recent development, $G_2$ metrics have been obtained
that make contact with the six-dimensional resolved and deformed
conifolds.  This work is described in detail in \cite{cglp12} (See
also \cite{cglp11,brand}).  We consider a generalization of the
original Ansatz \cite{brysal,gibpagpop} (\ref{g2old2}) for metrics of
cohomogeneity one with $S^3\times S^3$ principal orbits. The more
general Ansatz is given by
\bea
&&ds_7^2= 
dt^2 + c^2\, (\Sigma_3-\sigma_3)^2 + f^2\, (\Sigma_3 + g_3\,
\sigma_3)^2
\nn\\
&&+ a^2\, [(\Sigma_1 + g\, \sigma_1)^2 +
                          (\Sigma_2 + g\, \sigma_2)^2]
+ b^2\,  [(\Sigma_1 - g\, \sigma_1)^2 +
                          (\Sigma_2 - g\, \sigma_2)^2]\,,
\label{d7ans2}
\eea
where $\Sigma_i$ and $\sigma_i$ are again two sets of left-invariant
1-forms
of $SU(2)$, and the six coefficients $a$, $b$, $c$, $f$, $g$ and $g_3$ 
depend only on $t$.  In the orthonormal basis
\bea
&& e^0=dt\,,\quad  e^1= a\, (\Sigma_1 +  g\,
\sigma_1)\,,\quad  e^2= a\, (\Sigma_2 +  g\,
\sigma_2)\,,\quad  e^3 = c\, (\Sigma_3 - \sigma_3)\,,\nn\\
&& e^4 =  b\, (\Sigma_1 -  g\, \sigma_1)\,,\quad
 e^5 =  b\, (\Sigma_2 -  g\, \sigma_2)\,,\quad
 e^6 =  f\, (\Sigma_3 +  g_3\, \sigma_3)\,,\label{viel2}
\eea
there is a natural candidate for an invariant associative 3-form,
namely
\bea
\Phi &=&  e^0\wedge ( e^1\wedge  e^4 +
 e^2\wedge  e^5 +  e^3\wedge  e^6)
-( e^1\wedge  e^2- e^4\wedge  e^5)\wedge  e^3 \nn\\
&&+ ( e^1\wedge  e^5- e^2\wedge  e^4)\wedge  e^6
\,.\label{3form2}
\eea
Requiring the closure and co-closure of this 3-form gives a set of
first-order equations for $G_2$ holonomy \cite{cglp12},
\bea
\dot a &=& \fft{ c^2\, ( a^2 - b^2) +
[4 a^2\, ( a^2- b^2)-  c^2\, (5  a^2- b^2) - 4
a\,  b\,  c\,  f]\,  g^2}{16 a^2\,  b\,  c\,
g^2}\,,\nn\\
\dot b &=& -\, \fft{ c^2\, ( a^2- b^2) + [4  b^2\,
( a^2 - b^2) + c^2\, (5 b^2 -  a^2) - 4 a\,  b\,
 c\,  f]\, g^2}{16  a\,  b^2\,  c\,  g^2}\,,\nn\\
\dot c &=& \fft{ c^2 + ( c^2 -2 a^2 -2 b^2)\,
g^2}{4 a\,  b\,  g^2}\,,\label{5fo2}\\
\dot f &=& -\, \fft{( a^2- b^2)\, [ 4  a\,  b\,  f^2\,
 g^2 -  c\, (4 a\,  b\,  c +  a^2\,  f -
b^2\,  f)\, (1- g^2)]}{16  a^3\,  b^3\,  g^2}\,,\nn\\
\dot g &=& -\, \fft{ c\, (1- g^2)}{4 a\,  b\,  g}\,,\nn
\eea
together with an algebraic equation for $g_3$:
\be
 g_3 =  g^2 - \fft{ c\, ( a^2- b^2)
                 (1- g^2)}{2 a\,  b\,  f}\,.
\label{constr2}
\ee

   There are two combinationsof the equations (\ref{5fo2}) that can be
integrated explicitly, giving two invariants built out of the metric
functions.  These two constants are nothing but the coefficients in
front of the volume forms for the respective three-spheres in the
associated three-form: $\Phi=m\,
\sigma_1\wedge\sigma_2\wedge\sigma_3\, +\, n\,
\Sigma_1\wedge\Sigma_2\wedge\Sigma_3\,+\,\cdots $, which may be seen
to be constant by imposing closure of $\Phi$ (see \cite{brand}).
Ultimately, the system (\ref{5fo2}) can be reduced to a single
non-linear second-order differential equation.

   The general solution of the first-order equations (\ref{5fo2}) is
not known.  Of course the asymptotically conical $G_2$ metric
(\ref{g2old2}) is a solution. An explicit, singular, solution was found
in \cite{cglp11,brand}.  Another exact solution, found earlier in
\cite{bggg}, is
\bea
ds_7^2 &=& \fft{(r^2-\ell^2)}{(r^2-9\ell^2)}\, dr^2 
\ft1{12}(r-\ell)(r+3\ell)[(\Sigma_1-\sigma_1)^2+
(\Sigma_2-\sigma_2)^2] 
\nn\\
&&+\ft1{12}(r+\ell)(r-3\ell)[(\Sigma_1+\sigma_1)^2+
(\Sigma_2+\sigma_2)^2] \nn\\
&&+\ft19r^2\, (\Sigma_3-\sigma_3)^2
+ \ft49\ell^2\, \fft{r^2-9\ell^2}{r^2-\ell^2}\,
(\Sigma_3+\sigma_3)^2
\,.\label{g2met3}
\eea
The radial coordinate runs from an $S^3$ bolt at $r=3\ell$ to an
asymptotic region as $r$ approaches infinity.  The metric is
asymptotically locally conical, with the radius of the circle with
coordinate $(\psi+\wtd\psi)$ stabilising at infinity.  The metric is
closely analogous to an ALC Spin(7) metric on the $\R^4$ bundle over
$S^4$ that was found previously \cite{cglpspin7,cglpspin7m}.

   Although explicit solutions to the first-order system (\ref{5fo2})
are not in general known, it is nevertheless possible to study the
system by a combination of approximation and numerical methods.
Specifically, one can perform a Taylor expansion around the bolt at a
minimum radius where the $S^3\times S^3$ orbits degenerate, and use
this to set initial data just outside the bolt for a numerical
integration towards large radius.  The criterion for a complete
non-singular metric is that the metric functions should be
well-behaved at large distance, either growing linearly with distance
as in an AC metric, or else with one or more metric coefficients
stabilising to fixed values asymptotically, as in an ALC metric such
as (\ref{g2met3}).  This method is discussed in detail in
\cite{cglp8,cglp11,cglp12}, and it is established there that there
exist three families of non-singular ALC metrics, each with a
non-trivial parameter $\lambda$ that gives the size of a stabilising
circle at infinity relative to the size of the bolt at short distance.
The metrics, denoted by $\bB_7$, $\bD_7$ and $\wtd\bC_7$, have bolts
that are a round $S^3$, a squashed $S^3$ and $T^{p,q}=S^3\times
S^3/U(1)_{(p,q)}$ respectively, where $p/q=\sqrt{m/n}$ and $m,n$ are
the two explicit integration constants of the first-order system
(\ref{5fo2}) that we discussed previously.  The radius of the
stabilising circle ranges from zero at $\lambda=0$ to infinity at
$\lambda=\infty$.  As one takes the limit $\lambda\rightarrow 0$, the
ALC $G_2$ metric approaches the direct product of a six-dimensional
Ricci-flat K\"ahler metric and a vanishing circle.  This limit is
known mathematically as the Gromov-Hausdorff limit.

    The cases of most immediate interest are $\bB_7$ and $\bD_7$.
Their Gromov-Hausdorff limits are a vanishing circle times the
deformed conifold, and a vanishing circle times the resolved conifold,
respectively \cite{cglp11,cglp12}.  On the other hand, as $\lambda$
goes to infinity, they both approach the original AC metric of
\cite{brysal,gibpagpop}.  If, therefore, we begin with a solution
(Minkowski)$_4\times Y_7$ in M-theory, with $Y_7$ being a $\bB_7$ or
$\bD_7$ metric, then we can dimensionally reduce it on the circle that
stabilises at infinity, thereby obtaining a solution of the type IIA
string.  The radius of the M-theory circle, $R$, is related to the
string coupling constant $g_{\rm str}$ by $g_{\rm str}=R^{3/2}$.  This
means that taking the Gromov-Hausdorff limit in $\bB_7$ or $\bD_7$
corresponds to the weak-coupling limit in the type IIA string, and the
ten-dimensional solution becomes the product of (Minkowski)$_4$ with
the deformed or resolved conifold.  In the strong-coupling domain,
where $\lambda$ goes to infinity, these two ten-dimensional solutions
become unified via the $\bB_7$ and $\bD_7$ solutions in M-theory.

   A yet more general system of cohomogeneity one $G_2$ metrics with
$S^3\times S^3$ principal orbits was obtained recently in
\cite{cglp14}. The construction was based on an approach developed
recently by Hitchin \cite{hitch}, in which one starts from an Ansatz
for an associative 3-form, and derives first-order equations via a
system of Hamiltonian flow equations.  These first-order equations can
be shown to imply that a certain metric derived from the 3-form has
$G_2$ holonomy.  By applying this procedure to the case of $S^3\times
S^3$ principal orbits, it was shown in \cite{cglp14} that the metric
\crampest
\bea
&&ds_7^2 = dt^2 \\
&&+ {1\over {y_1}}[ (n\, x_1 + x_2\, x_3) \Sigma_1^2 +
 (m\, n + x_1^2 - x_2^2 - x_3^2)\Sigma_1\, \sigma_1 +
   (m\, x_1 + x_2\, x_3) \sigma_1^2]\nn\\
&&+{1\over {y_2}}[ (n\, x_2 + x_3\, x_1) \Sigma_2^2 +
 (m\, n + x_2^2 - x_3^2 - x_1^2)\, \Sigma_2 \sigma_2 +
   (m\, x_2 + x_3\, x_1)\sigma_2^2]\nn\\
&&+{1\over {y_3}}[ (n\, x_3 + x_1\, x_2) \Sigma_3^2 +
 (m\, n + x_3^2 - x_1^2 - x_2^2)\, \Sigma_3 \sigma_3 +
   (m\, x_3 + x_1\, x_2) \sigma_3^2]\nn\label{newmet}
\eea
\uncramp
has $G_2$ holonomy if the functions $x_i$ and $y_i$, which depend
only on $t$, satisfy the first-order Hamiltonian system of equations
\be
\dot x_1 = \sqrt{\fft{y_2\, y_3}{y_1}}\,,\qquad
\dot y_1 = \fft{m\, n\, x_1 + (m+n)\, x_2\, x_3
+ x_1\, (x_2^2+x_3^2-x_1^2)}{\sqrt{y_1\, y_2\, y_3}}\,,\label{newfo}
\ee
and cyclically for the 2 and 3 directions.  In addition, the conserved 
Hamiltonian must vanish, which implies that
\bea
&&4 y_1\, y_2\, y_3+
 m^2\, n^2 -2m\, n\, (x_1^2+x_2^2+x_3^2) - 4(m+n)\,
x_1\, x_2\, x_3\nn\\
&&+ x_1^4+x_2^4+x_3^4 - 2x_1^2\, x_2^2 - 2 x_2^2\, x_3^2 -2 x_3^2\,
x_1^2=0\,.
\eea

   The above first-order system encompasses all the previous cases as
specialisations.  In particular, the first-order system for the
metrics (\ref{d7ans2}) is obtained by making the specialisation
$x_1=x_2$, $y_1=y_2$.  If, instead, one sets $m=n=1$, the system
reduces to one studied in \cite{cglp5,bggg}. 

   In addition to these $SU(2)\times SU(2)$ invariant metrics with
principal orbits $S^3\times S^3$, one may take various
In\"on\"u-Wigner contractions to give metrics with principal orbits
$T^3\times S^3$, or other orbit types constructed from the possible
contractions of $SU(2)$ \cite{cglp14}.  In the particular case of
$T^3\times S^3$ orbits, the resulting first-order system is that of
\cite{zaslow}.

    We find that the general set of equations (\ref{newfo}) does not
seem to yield new classes of regular solutions, other than those
already classified \cite{cglp12} for the first-order system
(\ref{5fo2}).

\section{Conclusions and Open Avenues}
\label{Conclusions}

   In these lectures we have presented a summary of some recent
developments in the construction of regular $p$-brane configurations
with less than maximal supersymmetry. In particular, the method
involves the introduction of complete non-compact special holonomy
metrics and additional fluxes, supported by harmonic-forms in special
holonomy spaces, which modify the original $p$-brane solutions via
Chern-Simons ({\it transgression}) terms.

   The work led to a number of important {\it mathematical
developments} which we have also summarized.  Firstly, the
construction of harmonic forms for special holonomy spaces in diverse
dimensions was reviewed, and the explicit construction of harmonic
forms for Stenzel metrics was summarized.  Secondly, a construction of
new two-parameter Spin(7) holonomy spaces was discussed.  These have
the property that they interpolate asymptotically between a local
$S^1\times {\cal M}_7$, where the length of the circle is finite and
${\cal M}_7$ is the $G_2$ holonomy space with the topology of the
$S^2$ bundle over $S^4$, while at small distance they approach the
``old'' Spin(7) holonomy space with the topology of the chiral spin
bundle over $S^4$.

   These mathematical developments also led to a number of important
{\it physics implications}, relevant for the properties of the
resolved $p$-brane solutions.  In particular, the focus was on the
properties of resolved M2-branes with 8-dimensional special holonomy
transverse spaces, for example Stenzel, hyper-K\"ahler and Spin(7)
holonomy spaces, and the results for the fractional D2-branes with
three 7-dimensional $G_2$ holonomy transverse spaces.

   After the lectures were given, there was major progress in
constructing new $G_2$ holonomy spaces and studying the M-theory
dynamics on such spaces. We have summarized this progress in Section
\ref{New}, and in particular highlighted the classification of general
$G_2$ holonomy spaces with $S^3\times S^3$ principal orbits.

   Until recently, the emphasis has been on finding new $G_2$
manifolds that are complete and non-singular. However, M-theory
compactified on such spaces necessarily gives only Abelian and 
non-chiral ${\cal N}=1$ theories in four dimensions.   To obtain
non-Abelian chiral theories from M-theory, one needs to consider
compactifications on singular $G_2$ manifolds. One explicit
realisation of such an M-theory compactification has an interpretation
as an $S^1$ lift of Type IIA theory, compactified on an orientifold,
with intersecting D6-branes and O6 orientifold planes \cite{CSUII}.
Non-Abelian gauge fields arise at the locations of coincident branes,
and chiral matter arises at the intersections of
D6-branes. Interestingly, these constructions provide \cite{CSUI} the
first three-family supersymmetric standard-like models with
intersecting D6-branes.  The $S^1$ lift of these configurations
results in singular $G_2$ holonomy metrics in M-theory.  Co-dimension
four ADE-type singularities are associated with the location of the
coincident D6-branes, and co-dimension seven singularities are
associated with the location of the intersection of two D6-branes in
Type IIA theory \cite{CSUII,atwi,Witten,AcW,CSUIII}.

   Further analyses of co-dimension seven singularities of the $G_2$
holonomy spaces, leading to chiral matter, were given in
\cite{Witten,AcW,CSUIII} and subsequent work
\cite{Roiban,behrndt,Uranga02,LazarI,LazarII,bb,lust}.  It is expected
that there exists a wide classe of singular 7-manifolds with $G_2$
holonomy that yield non-Abelian ${\cal N}=1$ supersymmetric
four-dimensional theories with chiral matter.  The explicit
construction of such metrics would provide a starting point for
further studies of chiral M-theory dynamics.

    A recent study of an explicit class of singular $G_2$ holonomy
spaces was given in \cite{cglp17}.  These are cohomogeneity two
metrics foliated by twistor spaces, that is $S^2$ bundles over
self-dual Einstein four-dimensional manifolds $M_4$.  The 4-manifold
is chosen to be a self-dual Einstein space with orbifold
singularities.  An investigation of this construction was carried out
in \cite{behrndt}. In \cite{cglp17} the most general self-dual
Einstein metrics of triaxial Bianchi IX type, which have an $SU(2)$
isometry acting transitively on 3-dimensional orbits that are
(locally) $S^3$, were considered.

   Specialisation to biaxial solutions with positive cosmological
constant yields manifolds that are compact, but in general with
singularities.  The radial coordinate ranges over an interval that
terminates at endpoints where the $SU(2)$ principal orbits degenerate;
to a point (a NUT) at one end, and to a two-dimensional surface (a
bolt) that is (locally) $S^2$ at the other. Only for very special
values of the NUT parameter is the metric regular at both ends. In
general, however, one encounters singularities at both endpoints of
the radial coordinate.  In the generic case, a specific choice of the
period for the azimuthal angle allows the singularity at the $S^2$
bolt to be removed, but then the NUT has a co-dimension four orbifold
singularity. Alternatively, choosing the periodicity appropriate for
regularity at the NUT, there will be a co-dimension two singularity on
the $S^2$ bolt.  The associated seven-dimensional $G_2$ holonomy
spaces therefore have singularities of the same co-dimensions. The
co-dimension four NUT singularities may admit an M-theory
interpretation associated with the appearance of non-Abelian gauge
symmetries, and the circle reduction of M-theory on these $G_2$
holonomy spaces may have a Type IIA interpretation in terms of
coincident D6-branes \cite{behrndt}.  On the other hand, the
co-dimension two singularities at the bolts do not seem to have a
straightforward interpretation in M-theory.

\section*{Acknowledgement}

 Research is supported in part by DOE grant
DE-FG02-95ER40893, NSF grant No. PHY99-07949, Class of 1965
Endowed Term Chair and NATO Collaborative research grant 976951
(M.C.), in full by DOE grant DE-FG02-95ER40899 (H.L.) and in part
by DOE grant DE-FG03-95ER40917 (C.P.).

\end{document}